\documentclass[superscriptaddress,twocolumn,showpacs,preprintnumbers]{revtex4}
\usepackage[tbtags]{amsmath}
\usepackage{amsfonts}
\usepackage{mathrsfs}
\usepackage{amssymb}
\usepackage{graphicx}
\usepackage{epsfig,graphicx,times}
\usepackage{subfigure}
\usepackage{color}
\usepackage{amssymb}
\usepackage{amsmath}
\usepackage{graphicx}
\usepackage{dcolumn}
\usepackage{bm}
\usepackage{float}
\usepackage[abs]{overpic}

\begin{document}

\title{Tunable one-dimensional microwave emissions from cyclic-transition
three-level atoms}
\author{W. Z. Jia}
\affiliation{Quantum Optoelectronics Laboratory, School of Physics
and Technology, Southwest Jiaotong University, Chengdu 610031,
China}
\author{L. F. Wei}
\affiliation{Quantum Optoelectronics Laboratory, School of Physics
and Technology, Southwest Jiaotong University, Chengdu 610031,
China} \affiliation{State Key Laboratory of Optoelectronic Materials
and Technologies, School of Physics and Engineering, Sun Yat-Sen
University, Guangzhou 510275, China}
\author{Z. D. Wang}
\affiliation{Department of Physics and Center of Theoretical and Computational Physics,
The University of Hong Kong, Pokfulam Road, Hong Kong, China}
\date{\today}

\begin{abstract}
By strongly driving a cyclic-transition three-level artificial atom,
demonstrated by such as a flux-based superconducting circuit, we
show that coherent microwave signals can be excited along a coupled
one-dimensional transmission line. Typically, the intensity of the
generated microwave is tunable via properly adjusting the Rabi
frequencies of the applied strong-driving fields or introducing a
probe field with the same frequency. In practice, the system
proposed here could work as an on-chip quantum device with
controllable atom-photon interaction to implement a total-reflecting
mirror or switch for the propagating probe field.
\end{abstract}

\maketitle




\bigskip




\bigskip

\textit{Introduction.---}Superconducting quantum circuits (SQCs) can
be regarded as artificial atoms (AAs) with quantized energy
levels~\cite{martinis88}. Quantum mechanical behaviors in these AAs,
such as spectroscopy~\cite{spe1,spe2,spe3}, Rabi
oscillations~\cite{Rabi1,Rabi2}, and so forth, have already been
demonstrated experimentally. Additionally, strong coupling between
an AA and a single-mode microwave field in a high-Q resonator can
realize a macroscopic analog of the cavity quantum electrodynamics
(QED), known as circuit QED~\cite{cQED1,cQED2}. While, when such an
atom interacts with waves propagating freely along an open 1D
transmission line (TL), the situation is qualitatively different. In
this case, the system should be described by the continuous
electromagnetic fields being scattered by a point-like AA~\cite
{Shen2005,NECscience2010,EITinAA,OnChipAmplifer}. Differing from the
usual 3D scattering process, here the problem of spatial-mode
mismatch between the incident and scattered waves can be effectively
overcome as these waves are confined in the 1D
space~\cite{Shen2005,NECscience2010}. This also provides a way to
demonstrate the strong interactions between the 1D microwave fields
and an AA based on flux-biased superconducting
circuit~\cite{NECscience2010,EITinAA,OnChipAmplifer}. Specifically,
by virtue of the tunability, controllability, and strong atom-photon
coupling, these on-chip macroscopic quantum devices could be
utilized to reproduce certain typical quantum optical phenomena,
e.g., resonance fluorescence~\cite{NECscience2010},
electromagnetically induced transparency (EIT)~\cite{EITinAA}, and
ultimate on-chip amplifier~\cite{OnChipAmplifer}, etc..

In this paper, we discuss the feasibility of generating microwave
with specific frequency along an open 1D TL by manipulating the
strong-field-induced coherence in a coupled three-level $\Delta
$-type AA~\cite{martinis88,DeltaAtom1,DeltaAtom2}. The produced
microwave emission can be controlled by introducing another weak
probe field with the same frequency. Our results display some
special 1D quantum optical phenomena in microwave domain and provide
the potential applications in on-chip photonic quantum information
processings with SQCs.

\begin{figure}[b]
\includegraphics[width=0.25\textwidth]{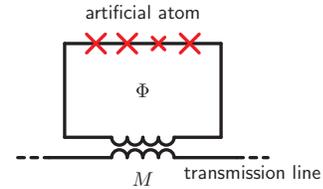}
\caption{(Color online) Circuit diagram~\protect\cite{NECscience2010} of an
artificial atom (generated by four-junction flux qubit geometry) inductively
coupled to a 1D long transmission line.}
\label{circuit}
\end{figure}

\textit{Tunable microwave emissions from a $\Delta$-type artificial
atom.---}We consider a system schematized in Fig.~\ref{circuit},
where a superconducting loop interrupted by four Josephson junctions
couples to an open 1D TL through the loop-line mutual inductance
$M$. The loop can be looked on as a multilevel AA, whose energy
levels and transition elements can be tuned by adjusting the
external magnetic flux $\Phi $. At some working
points~\cite{OnChipAmplifer}, the desirable three-level $\Delta
$-type transition configurations can be
realized~\cite{DeltaAtom1,DeltaAtom2}. Here we assume $\left\vert
i\right\rangle $ ($i=1,2,3$) and $\omega_{i}$ are the three selected
levels and their eigenfrequenies of the $\Delta$-type AA. Three
coherent driving fields $\frac{1}{2}I_{ij}e^{ik_{ij}x-\nu_{ij}t}+
\mathrm{c.c.}$ ($i,j=1,2,3$, $i>j$), propagating along the open 1D
TL with the wave vectors $k_{ij}$ and frequencies $\nu _{ij}$, are
applied to couple the three possible transition-channels of the
point-like AA located at $x=0$. The Hamiltonian of this system can
be written as
$\hat{H}=\hbar \sum_{i=1}^{3}\omega _{i}\hat{\sigma}_{ii}-\left(
\hbar /2\right)
\sum_{i>j=1}^{3}\left( \Omega _{ij}e^{-i\nu _{ij}t}\hat{\sigma}_{ij}+%
\mathrm{H.c.}\right) $.
Here $\hat{\sigma}_{ij}=\left\vert i\right\rangle \left\langle
j\right\vert $ are the atomic projection or transition operators,
and $\Omega _{ij}=\phi _{ij}I_{ij}/\hbar$ are the Rabi frequencies.
The dipole moment matrix elements can be written as $\phi
_{ij}=t_{ij}MI_{P}$, where $t_{ij}$ ($=t_{ji}$) are the
dimensionless matrix elements, $M$ is the line-atom mutual
inductance, and $I_{P}$ is the amplitude of the persistent current
in the loop.

\begin{figure}[t]
\includegraphics[width=0.5\textwidth]{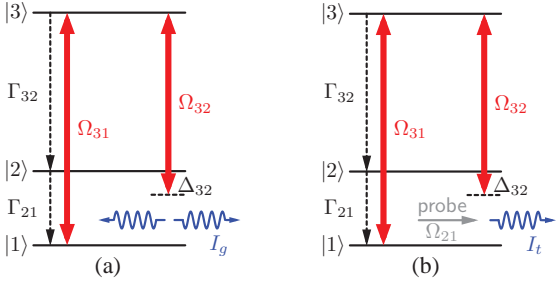} \put(-200,-5){(a)}
\put(-80,-5){(b)}
\caption{{}(Color online) Schematics of the driven three-level artificial
atom: (a) Strong fields with Rabi frequencies $\Omega _{31}$ and $\Omega
_{32}$ are applied to produce microwave emission between the levels $%
\left\vert 1\right\rangle $ and $\left\vert 2\right\rangle $. (b) A weak
probe field with Rabi frequency $\Omega _{21}$ is additionally applied to
control the microwave emission.}
\label{EL}
\end{figure}

In the following, we will show that two strong-driving fields
$\Omega _{31}$ and $\Omega _{32}$ can induce an effective magnetic
flux $\phi $ with oscillating frequency $\nu _{21}=\nu _{31}-\nu
_{32}$ in the AA-loop. And consequently, the AA as a point-like
source, will emit microwaves with frequency $\nu _{21}$ in the TL.
Specifically, we assume $\Omega _{31}$ is resonantly applied between
the levels $\left\vert 1\right\rangle $ and $\left\vert
3\right\rangle $, and $\Omega _{32}$ applied between the levels
$\left\vert 2\right\rangle $ and $\left\vert 3\right\rangle $ with a
detuning $\Delta _{32}=\nu _{32}-\omega _{3}+\omega _{2}$, as shown
in Fig.~\ref{EL}(a). In the interaction picture, the Hamiltonian of
this driven three-level system reads
$\hat{H}_{I}=\hbar \Delta _{32}\hat{\sigma}_{22}-\frac{\hbar
}{2}\left( \Omega _{31}\hat{\sigma}_{31}+\Omega
_{32}\hat{\sigma}_{32}+\mathrm{H.c.}\right)$,
and its corresponding dynamics can be described by the master
equation
\begin{equation}
\dot{\rho}=-\left( i/\hbar \right)[ \hat{H}_{I},\rho] +L[\rho]
\label{ME}
\end{equation}
with $\rho =\sum_{ij}\rho _{ij}\hat{\sigma}_{ij}$ being the density matrix.
The Lindblad term is defined by
$L\left[ \rho \right] =\Gamma _{31}\rho _{33}\left(
\hat{\sigma}_{11}-\hat{\sigma}_{33}\right) +\Gamma _{32}\rho
_{33}\left( \hat{\sigma}_{22}-\hat{\sigma}_{33}\right) +\Gamma
_{21}\rho _{22}\left( \hat{\sigma}_{11}-\hat{\sigma}_{22}\right)
-\sum_{i\neq j}\gamma _{ij}\rho _{ij}\hat{\sigma}_{ij}$.

Under the usual steady-state condition, the density matrix element $\rho
_{21}$ (which describes the coherence between $\left\vert 1\right\rangle $
and $\left\vert 2\right\rangle $) can be written as
\begin{equation}
\rho _{21}=\frac{\Omega _{31}\Omega _{32}^{\ast }\left[ \lambda _{23}\left(
\rho _{33}-\rho _{11}\right) +\gamma _{13}\left( \rho _{33}-\rho
_{22}\right) \right] }{\left\vert \Omega _{31}\right\vert ^{2}\gamma
_{13}+\lambda _{23}\left( \left\vert \Omega _{32}\right\vert ^{2}+4\gamma
_{13}\lambda _{21}\right) }  \label{coherent1a}
\end{equation}
with $\lambda _{21}=\gamma _{12}-i\Delta $, $\lambda _{23}=\gamma
_{23}-i\Delta $. Here $\Delta =-\Delta _{32}$ can be defined as the
detuning of the induced wave. Based on the experimental
detections~\cite{OnChipAmplifer}, we have $\Gamma _{32}\gg \Gamma
_{21}$, $\Gamma _{31}$, and $\gamma _{13}\approx \Gamma _{32}/2$,
$\gamma _{23}\approx \Gamma _{32}/2 $. As a consequence, the steady
populations at resonant point $\Delta =0$ read
$\rho _{11}=( 2\gamma
_{12}+\Gamma _{32}) \left\vert \Omega _{32}\right\vert ^{2}/A$,
$\rho _{22}=[ 2\gamma _{12}\left\vert \Omega _{32}\right\vert
^{2}+\Gamma _{32}( \left\vert \Omega _{31}\right\vert ^{2}+2\gamma
_{12}\Gamma _{32})] /A$,
and
$\rho _{33}=2\gamma _{12}\left\vert \Omega _{32}\right\vert
^{2}/A$, where $A=( 6\gamma _{12}+\Gamma _{32}) \left\vert \Omega
_{32}\right\vert ^{2}+\Gamma _{32}( \left\vert \Omega
_{31}\right\vert ^{2}+2\gamma _{12}\Gamma _{32})$.
Thus $\rho_{21}$ (at resonant point $\Delta =0$) can be simplified
as
\begin{equation}
\rho _{21}=-\Gamma _{32}\Omega _{31}\Omega _{32}^{\ast }/A.
\label{coherent1b}
\end{equation}
This implies that an additional magnetic flux threading the AA-loop
$\phi \left( t\right) =\mathrm{Tr}(\hat{\phi}\left( t\right) \rho
)=\phi _{12}\rho _{21}e^{-i\nu _{21}t}+\mathrm{c.c.}$ can be
induced,
where $\hat{\phi}\left( t\right) =\phi _{12}
\hat{\sigma}_{12}e^{-i\nu _{21}t}+\mathrm{H.c.}$ is the related
atomic dipole-moment operator.  Note that this flux is originated
from the nonzero dipole-transition elements $\phi _{ij}$ between any
pair of levels, which can only be realized in the present
$\Delta$-type AA with the broken parity-symmetry.

Certainly, this induced oscillating flux in the AA-loop could produce
mutually a microwave current $I_{g}\left( x,t\right) $ along the TL, which
satisfies the relevant 1D wave equation:
\begin{equation}
\left( \partial _{xx}-v^{-2}\partial _{tt}\right) I_{g}\left(
x,t\right) =c\delta \left( x\right) \partial _{tt}\phi \left(
t\right) ,  \label{1D WE}
\end{equation}
where $v=1/\sqrt{lc}$ is the phase velocity ($l$ and $c$ are
inductance and capacitance per unit length, respectively) and the
dispersion relation is $\nu _{21}=vk_{21}$. The above equation
describes a point-like AA located at $x=0$ with oscillating flux
$\phi \left( t\right) $ emitting microwave $I_{g}\left( x,t\right) $
in the 1D TL. With the relation $\Gamma _{ij}=( \hbar \omega
_{ij}\phi _{ij}^{2}) /( \hbar ^{2}Z) $ (with $Z=\sqrt{l/c}$ being
the line impedance)~\cite{NECscience2010}, the solution of
Eq.~\eqref{1D WE} can be expressed as $I_{g}\left( x,t\right)
=\frac{1}{2}I_{g}e^{ik_{21}\left\vert x\right\vert -i\nu
_{21}t}+\mathrm{c.c.}$ (representing the induced waves propagating
in both directions)\ with the complex amplitude $I_{g}=iJ\rho
_{21}$, and $J=\sqrt{\hbar \omega _{21}\Gamma _{21}/Z}$. Thus the
intensity of the induced microwave current reads:
$\left\vert I_{g}\right\vert =J\left\vert \rho _{21}\right\vert $.
Specifically, at the resonant point, $\left\vert I_{g}\right\vert $
is determined by
\begin{equation}
\left\vert I_{g}\right\vert =J\Gamma _{32}\left\vert \Omega _{31}\right\vert
\left\vert \Omega _{32}\right\vert /A,  \label{GWI1b}
\end{equation}
and could be controlled via adjusting the values of the applied Rabi
frequencies. For example, if $\left\vert \Omega _{31}\right\vert $
and $\left\vert \Omega _{32}\right\vert $ are modulated
synchronously, i.e., $\left\vert \Omega _{31}\right\vert =\left\vert
\Omega _{32}\right\vert =\Omega $, then $\left\vert
I_{g}\right\vert$ is monotonically increasing with $\Omega$,
approaching $J\Gamma _{32}/\left( 6\gamma _{12}+2\Gamma _{32}\right)
$ for sufficiently large $\Omega $. Typically, the upper limit of
the produced microwave current $\left\vert I_{g}\right\vert
_{\mathrm{max}}=J/2$ is achieved for $\gamma _{12}\ll \Gamma_{32}$.

The above microwave emission could be further controlled by applying
a probe field $I_{21}(x,t)=\frac{1}{2}I_{21}e^{ik_{21}x-i\nu
_{21}t}+\mathrm{c.c.}$ to couple the levels $\left\vert
1\right\rangle $ and $\left\vert 2\right\rangle $, as shown in
Fig.~\ref{EL}(b). At the resonant point and under the weak probe
approximation, the according steady-state coherence term is
\begin{equation}
\rho _{21}^{\prime }=-\Gamma _{32}\Omega _{31}\Omega _{32}^{\ast }/A+\Omega
_{21}B  \label{Coherent2}
\end{equation}
with
\begin{equation*}
B=\frac{i\Gamma _{32}\left[ \Gamma _{32}\left( \left\vert \Omega
_{31}\right\vert ^{2}-\left\vert \Omega _{32}\right\vert ^{2}\right)
+2\gamma _{12}\left( \Gamma _{32}^{2}-\left\vert \Omega _{32}\right\vert
^{2}\right) \right] }{A\left( \left\vert \Omega _{31}\right\vert
^{2}+\left\vert \Omega _{32}\right\vert ^{2}+2\gamma _{12}\Gamma
_{32}\right) }.
\end{equation*}

Clearly, the net microwave propagating along the 1D TL reads
$I_{t}\left(
x,t\right) =\frac{1}{2}\left( I_{21}e^{ik_{21}x}+iJ\rho _{21}^{\prime
}e^{ik_{21}\left\vert x\right\vert }\right) e^{-i\nu _{21}t}+\mathrm{c.c.}$.
And the amplitude of resulting current at $x>0$ can be expressed as
$I_{t}=I_{21}+iJ\rho _{21}^{\prime}$, namely, a superposition of the
incident probe and the wave mainly induced by the strong driving
fields. Utilizing the expression for $\Omega _{ij}$ and $\Gamma
_{ij}$, the intensity of the total microwave current $\left\vert
I_{t}\right\vert $ can be written as
$\left\vert I_{t}\right\vert
=J\left\vert \Omega _{21}/\Gamma _{21}+i\rho _{21}^{\prime
}\right\vert $.
For strong driving fields, the second term in Eq.~\eqref{Coherent2}
is much smaller than the first term in Eq.~\eqref{Coherent2} and
$\Omega _{21}/\Gamma_{21}$~\cite{sup}, and can then be neglected
when calculating $\left\vert I_{t}\right\vert$. Note that this
approximation is optimal if choosing $\left\vert \Omega
_{31}\right\vert =\left\vert \Omega_{32}\right\vert =\Gamma _{32}$,
resulting in vanishing of the second term in Eq.~\eqref{Coherent2}.
Consequently, at resonant point, $\left\vert I_{t}\right\vert$ can
be rewritten as
\begin{equation}
\left\vert I_{t}\right\vert =\left( J\left\vert \Omega _{21}\right\vert
/\Gamma _{21}\right) \left( 1+\alpha ^{2}-2\alpha \sin \Theta \right) ^{%
\frac{1}{2}},  \label{GWI2a}
\end{equation}
with
$\alpha =\Gamma _{21}\Gamma _{32}\left\vert \Omega _{31}\right\vert
\left\vert \Omega _{32}\right\vert /\left( A\left\vert \Omega
_{21}\right\vert \right) $, and $\Theta =\theta _{21}+\theta _{32}-\theta
_{31}$ (where $\theta _{ij}$ is the phase factor of $\Omega _{ij}$).

The above calculations indicate that: (1) For the fixed complex Rabi
frequencies $\Omega _{31}$ and $\Omega _{32}$, $\left\vert
I_{t}\right\vert $ can still be modulated by changing the complex
Rabi frequency of the probe $\Omega _{21}=\left\vert \Omega
_{21}\right\vert e^{i\theta _{21}}$. Particularly, when $\Theta =\pi
/2$, $\alpha =1$, the induced microwave current at $x>0$ is totally
switched off (i.e., $\left\vert I_{t}\right\vert =0$). Certainly, in
this case the induced microwave current at $x<0$ remains unchanged
and thus the relevant microwave emission from the AA is
\textit{unidirectional}. (2) Alternatively, if the probe
$\Omega_{21}$ is treated as a signal, the above strong-field-driven
AA can be regarded as a total reflection mirror for the probe field.
Originally, in the absence of the strong-driving fields, a weak
probe field propagating along the 1D TL can be partially absorbed
(reflected) by resonantly interacting with the AA. The reflection
amplitude could be expressed as $\Gamma_{21}/\left( 2\gamma
_{12}\right)$, if the atom-line coupling efficiency is
unit~\cite{NECscience2010}. Therefore, only when the pure dephasing
$\Gamma _{12}^{\varphi }$ ($=\gamma _{12}-\Gamma _{21}/2$) is
negligible, the AA can work as the desirable reflection mirror. Our
calculations show that, this shortcoming could be overcome by
appropriately applying the two strong-driving fields $\Omega _{31}$
and $\Omega _{32}$ satisfying $\Theta =\pi /2$; $\alpha =1$. Under
this condition, despite of remarkable pure dephasing (note that
$\alpha $ is a function of $A\left( \gamma _{12}\right) $), the
incident probe field can be \textit{completely} canceled out because
of its destructive interference with the strong-field-induced wave.
Consequently, an interesting strong-field-induced absorption
(reflection) phenomenon can be realized.

\begin{figure}[t]
\includegraphics[width=0.5\textwidth]{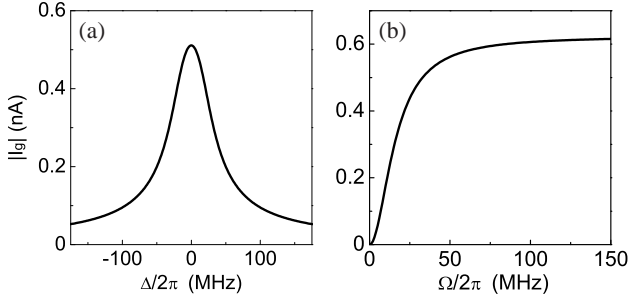} \put(-223,100){(a)}
\put(-109.5,100){(b)} \caption{(a) The microwave emission spectrum
induced by strong driving fields $\Omega _{31}$ and $\Omega _{32}$.
(b) The intensity of produced microwave current as a function of
$\Omega $ ($\left\vert \Omega _{31}\right\vert =\left\vert \Omega
_{32}\right\vert =\Omega $) at $\Delta =0 $.} \label{GW}
\end{figure}

\begin{figure}[t]
\begin{minipage}{0.26\textwidth}
\centering
\includegraphics[height=4cm]{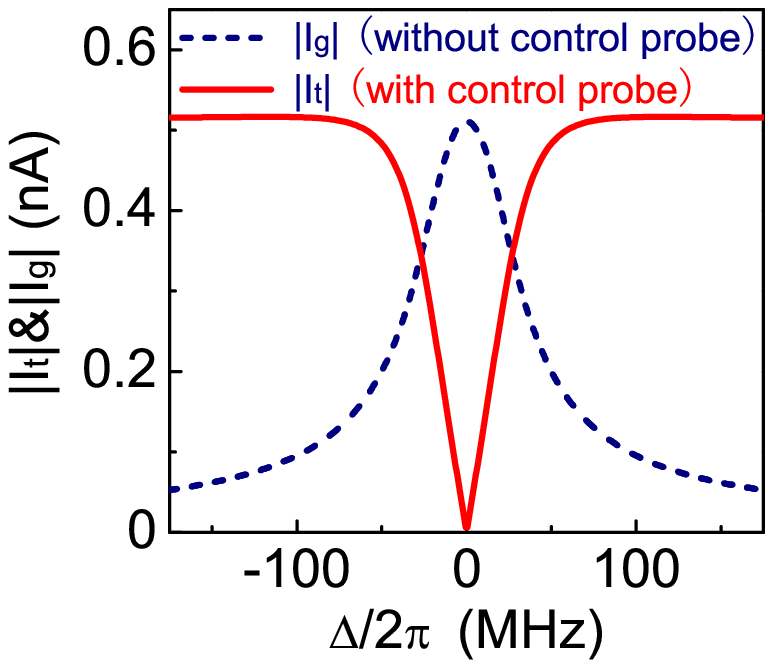}
\end{minipage}
\begin{minipage}{0.21\textwidth}
\centering
\includegraphics[height=3.6cm]{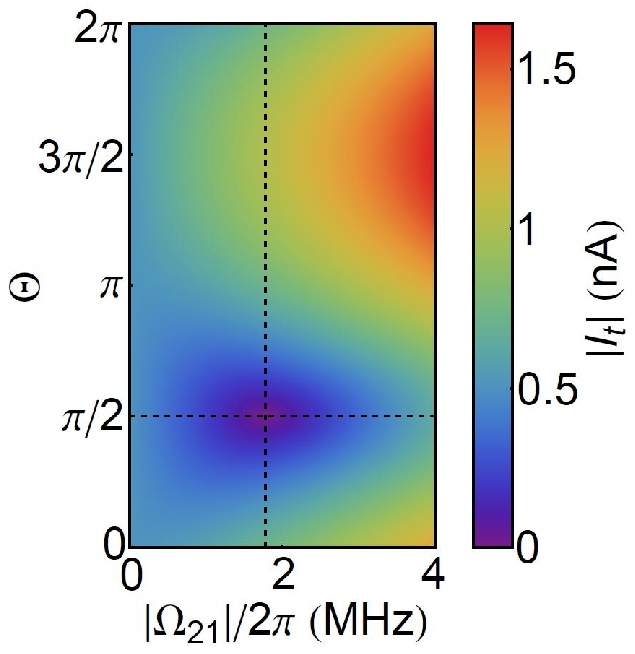}
\end{minipage}
\put(-205,24){(a)} \put(-78,42){(b)} \newline \centering
\caption{(Color online) Switch-off the produced microwave emission
by a probe field. (a) Induced microwave emission spectrum without
control probe (dashed line), which can be switched off at resonant
point by a control probe $\Omega _{21}$ (solid line). (b) The
intensity plot displays the
optimal working point for the probe to switch off the microwave emission: $%
\Theta =\protect\pi /2$, $\left\vert \Omega _{21}\right\vert /2\protect\pi %
=1.78$MHz. } \label{GWwithP}
\end{figure}

\begin{figure}[t]
\centering
\includegraphics[width=0.5\textwidth]{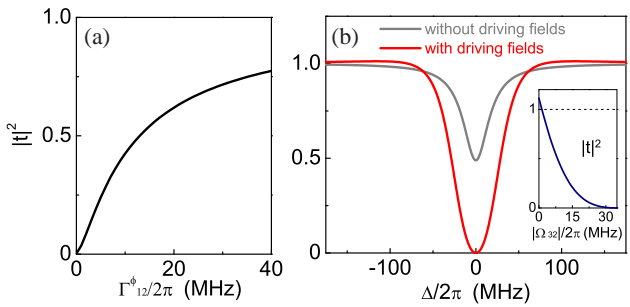} \put(-224,102){(a)}
\put(-130.5,102){(b)}
\caption{(Color online) (a) When no driving fields applied, the influence of
pure dephasing $\Gamma _{12}^{\protect\varphi }$ on the power transmission
coefficient $\left\vert t\right\vert ^{2}$ for a resonantly incident weak
probe with $\left\vert \Omega _{21}\right\vert /2\protect\pi =1.78$MHz. (b)
The power transmission coefficient $\left\vert t\right\vert ^{2}$ as a
function of probe detuning, displaying partial absorption when no driving
fields applied and strong-field-induced absorption, respectively. The inset
shows single AA quantum switch for resonant probe signal, realized\ by
fixing $\left\vert \Omega _{31}\right\vert $\ and modulating $\left\vert
\Omega _{32}\right\vert $.}
\label{CReflection}
\end{figure}

\textit{Experimental feasibility.---} Immediately, the phenomena
predicted above could be verified with the recent experimental
device~\cite{NECscience2010,EITinAA,OnChipAmplifer}, i.e., an open
1D TL couples to an AA based on superconducting flux qubit geometry.
For the four-junction flux qubit proposed there, by tuning the flux
bias $\delta \Phi =\Phi-\Phi _{0}/2$ (with $\Phi _{0}$ being the
flux quantum) slightly apart from the point $\delta \Phi =0$, a
cyclic-transition structure can be realized with the lowest three
levels. Typically, if the flux bias is set as $\delta \Phi /\Phi
_{0}=3.5\times 10^{-3}$, the corresponding experimental parameters
of AA are $\Gamma _{32}/2\pi =35$MHz, $\Gamma _{21}/2\pi =11$MHz,
$\gamma _{12}/2\pi =18$MHz, $\omega _{32}/2\pi =24.15$GHz, $\omega
_{21}/2\pi =10.96$GHz~\cite{OnChipAmplifer}. Also, the
characteristic impedance of the coplanar TL is $Z\approx 50\Omega$
~\cite{EITinAA}. Furthermore, we assume that the Rabi frequencies of
the applied two strong-driving fields could be set as $\left\vert
\Omega _{31}\right\vert /2\pi =\left\vert \Omega _{32}\right\vert
/2\pi =35$MHz.
With these parameters, numerical simulations are carried out, and
the results are presented in Figs.~\ref{GW}-\ref{CReflection}.

We plot the induced microwave emission spectrum in Fig.~\ref{GW}(a).
It is seen that, around the resonant point $\Delta =0$, an obvious
emission peak appears with maximal amplitude $\left\vert
I_{g}\right\vert \approx 0.51$nA, which is in good agreement with
the above theoretical estimation utilizing Eq.~\eqref{GWI1b}. In
addition, Fig.~\ref{GW}(b) shows the influence of driving fields on
$\left\vert I_{g}\right\vert$ (Here, $\left\vert \Omega
_{31}\right\vert =\left\vert \Omega _{32}\right\vert =\Omega$). We
can see that a remarkable microwave current can be obtained when the
driving fields are comparable with $\Gamma _{32}$. It reaches
$0.62$nA, if the control fields are strong enough. This is in good
accordance with the theoretical prediction ($J\Gamma _{32}/\left(
6\gamma _{12}+2\Gamma _{32}\right) $) given above. To attain a more
remarkable induced current, one can enhance either $\left\vert \rho
_{21}\right\vert $ or $J$. On one hand, a smaller decoherence rate
$\gamma _{12}$ may lead to a larger $\left\vert \rho
_{21}\right\vert $. Note that in the limit of $\gamma _{12}\ll
\Gamma _{32}$, the upper limit of induced current is $J/2=1.58$nA.
On the other hand, as $J=\sqrt{\hbar \omega _{21}\Gamma _{21}/Z}$,
one can enlarge $J$ by choosing sample with smaller line impedance
$Z$ and larger loop-line mutual inductance $M$.

Fig.~\ref{GWwithP}(a) shows that, when a probe field with $\Theta
=\pi /2$, $\left\vert \Omega _{21}\right\vert /2\pi =1.78$MHz
(satisfying $\alpha =1$) is applied, the original induced microwave
emission peak (dashed line, repeating the spectrum in Fig.~\ref{GW}
(a)), will be changed into a dip with zero emission at resonant
point (solid line). This indicates that the total suppression for
the emission current at $x>0$, which is in accordance with the above
theoretical estimation. Fig.~\ref{GWwithP}(b) displays $\left\vert
I_{t}\right\vert$ as a function of $\left\vert \Omega
_{21}\right\vert$ and $\Theta$ (The detuning is fixed at $\Delta
=0$). This verifies that the probe parameters used in
Fig.~\ref{GWwithP}(a) are optimal for implementing the desirable
switch-off operation.

By setting $\Omega_{31}=\Omega_{32}=0$, the power transmission
coefficient $\left\vert t\right\vert ^{2}$ ($t=I_{t}/I_{21}$) as a
function of pure dephasing $\Gamma _{12}^{\varphi }$ (other atom
parameters are the same as previous discussion), for a resonantly
incident weak probe with $\left\vert \Omega _{21}\right\vert /2\pi
=1.78$ MHz, is shown in Fig.~\ref{CReflection}(a). One can see that,
only in the ideal case of $\Gamma _{12}^{\varphi }=0$ the atom can
be regarded as a total reflection mirror with $\left\vert
t\right\vert ^{2}$ $=0$. However, when the pure dephasing is not
negligible, only partial power can be reflected by the atom.
Typically, for a probe with $\left\vert \Omega _{21}\right\vert
/2\pi =1.78$ MHz interacting with an AA with pure dephasing $\Gamma
_{12}^{\varphi }/2\pi =12.5$MHz (gotten from AA parameters in
Ref.~\cite{OnChipAmplifer}), at resonant point the according power
transmission coefficient $\left\vert t\right\vert ^{2}$ is only
about $0.49$ (gray line in Fig.~\ref{CReflection}(b)). When the two
strong fields with $\left\vert \Omega _{31}\right\vert /2\pi
=\left\vert \Omega _{32}\right\vert /2\pi =35$MHz (satisfying
$\alpha =1$), $\Theta =\pi /2$ are applied, at the resonant point,
the power extinction for the same probe is reaching the ideal case
of 100\% (red line in Fig.~\ref{CReflection}(b)), displaying the
phenomenon of strong-field-induced absorption (single atom total
reflection mirror). Moreover, the inset in Fig.~\ref{CReflection}(b)
shows by fixing other parameters and decreasing $\left\vert \Omega
_{32}\right\vert $ from $2\pi \times 35$MHz to about $2\pi \times
1.3$MHz, the resonant probe signal can be tuned from completely
reflection to totally transmission, demonstrating a perfect single
AA switch for propagating probe. Also, the inset verifies when
$\left\vert \Omega _{32}\right\vert =0$, the AA will work as an
on-chip amplifier~\cite{OnChipAmplifer}.

\textit{Conclusions.---}In summary, we have proposed an approach to
realize strong-field-induced microwave emission by a single AA
coupled to 1D open space of a TL. Moreover, we have shown that this
kind of induced microwave emission is controllable and tunable by
another weak probe. Although various quantum optical effects based
on multilevel structure of superconducting AAs have been
demonstrated in a series of recent
experiments~\cite{EITinAA,ATinSQC1,ATinSQC2,CPTinSQC}, our results
show certain novel effects in the 1D quantum optics at microwave
regime. On the other hand, the quantum devices with these effects
may have interesting applications in photonic quantum information
processing utilizing SQCs. Substantially, the generation of induced
microwave emission is a frequency conversion process. This indicates
that the proposed device may be used to connect superconducting
quantum devices operating at different frequencies in the future
SQCs. Finally, the device can also be used as tunable coherent
microwave source, or total reflection mirror (single AA quantum
switch) controlling the propagation of 1D probe field.

\begin{acknowledgments}
The project was supported in part by National Natural Science Foundation of
China under Grant Nos. 10874142, 90921010, and the National Fundamental
Research Program of China through Grant No. 2010CB923104.
\end{acknowledgments}



\begin{thebibliography}{99}
\bibitem{martinis88} J. Clark, A. N. Cleland, M. H. Devoret, D. Esteve, and
J. M. Martinis, Science \textbf{239}, 992 (1988).

\bibitem{spe1} J. R. Friedman, V. Patel, W. Chen, S. K. Tolpygo, and J. E.
Lukens, Nature (London) \textbf{406}, 43 (2000).

\bibitem{spe2} C. H. van der Wal, A. C. J. ter Haar, F. K. Wilhelm, R. N.
Schouten, C. J. P. M. Harmans, T. P. Orlando, S. Lloyd, and J. E. Mooij,
Science \textbf{290}, 773 (2000);

\bibitem{spe3} A. J. Berkley, H. Xu, R. C. Ramos, M. A. Gubrud, F. W.
Strauch, P. R. Johnson, J. R. Anderson, A. J. Dragt, C. J. Lobb, and F. C.
Wellstood, Science \textbf{300}, 1548 (2003).

\bibitem{Rabi1} D. Vion, A. Aassime, A. Cottet, P. Joyez, H. Pothier, C.
Urbina, D. Esteve, and M. H. Devoret, Science \textbf{296}, 886 (2002).

\bibitem{Rabi2} Chiorescu, Y. Nakamura, C. J. P. M. Harmans, and J. E.
Mooij, Science \textbf{299}, 1869 (2003).

\bibitem{cQED1} A. Wallraff, D. I. Schuster, A. Blais, L. Frunzio, R. S.
Huang, J. Majer, S. Kumar, S. M. Girvin, and R. J. Schoelkopf, Nature
(London) \textbf{431}, 162 (2004);

\bibitem{cQED2} A. Blais, R. S. Huang, A. Wallraff, S. M. Girvin, and R. J.
Schoelkopf Phys. Rev. A \textbf{69}, 062320 (2004).

\bibitem{Shen2005} J. T. Shen and S. Fan, Phys. Rev. Lett., \textbf{95},
213001 (2005).

\bibitem{NECscience2010} O. Astafiev, A. M. Zagoskin, A. A. Abdumalikov Jr.,
Y. A. Pashkin, T. Yamamoto, K. Inomata, Y. Nakamura and J. S. Tsai, Science
\textbf{327}, 840 (2010).

\bibitem{EITinAA} A. A. Abdumalikov, Jr., O. Astafiev, A. M. Zagoskin, Yu.
A. Pashkin, Y. Nakamura, and J. S. Tsai, Phys. Rev. Lett. \textbf{104},
193601 (2010).

\bibitem{OnChipAmplifer} O. Astafiev, A. A. Abdumalikov, Jr., A. M.
Zagoskin, Yu. A. Pashkin, Y. Nakamura, and J. S. Tsai, Phys. Rev. Lett.
\textbf{104}, 183603 (2010).

\bibitem{DeltaAtom1} Yu-xi Liu, J. Q. You, L. F. Wei, C. P. Sun, and F.
Nori, Phys. Rev. Lett. \textbf{95}, 087001 (2005).

\bibitem{DeltaAtom2} L. F. Wei, J. R. Johansson, L. X. Cen, S. Ashhab, and Franco
Nori, Phys. Rev. Lett. \textbf{100}, 113601 (2008); W. Z. Jia and L.
F. Wei, Phys. Rev. A \textbf{82}, 013808 (2010).

\bibitem{sup} With the experimental parameters in Ref. \cite{OnChipAmplifer}, numerical
calculation shows that if $\left\vert \Omega _{21}\right\vert
=\Gamma
_{21}/5$, $\left\vert \Omega _{31}\right\vert \geqslant \Gamma _{32}$, $%
\left\vert \Omega _{32}\right\vert \geqslant \Gamma _{32}$, then
$\left\vert (\Omega _{21}B)/\left( \Gamma _{32}\Omega _{31}\Omega
_{32}^{\ast }/A\right) \right\vert <0.02$, $\left\vert (\Omega
_{21}B)/\left( \Omega _{21}/\Gamma _{21}\right) \right\vert <0.015$.

\bibitem{ATinSQC1} M. Baur, S. Filipp, R. Bianchetti, J. M. Fink, M. Goppl,
L. Steffen, P. J. Leek, A. Blais, and A. Wallraff, Phys. Rev. Lett. \textbf{%
102}, 243602 (2009).

\bibitem{ATinSQC2} M. A. Sillanpa, Jian Li, K. Cicak, F. Altomare, J. I.
Park, R. W. Simmonds, G. S. Paraoanu, and P. J. Hakonen, Phys. Rev. Lett.
\textbf{103}, 193601 (2009).

\bibitem{CPTinSQC} W. R. Kelly, Z. Dutton, J. Schlafer, B. Mookerji, T. A.
Ohki, J. S. Kline and D. P. Pappas, Phys. Rev. Lett. \textbf{104}, 163601
(2010).
\end{thebibliography}
\end{document}